\documentclass[aps,prl,twocolumn,preprintnumbers,groupedaddress]{revtex4}

\usepackage{amssymb}
\usepackage{epsfig} 
\usepackage{amsmath,color} 
\usepackage{graphicx} 
\usepackage{ulem}

\usepackage{wrapfig}

\newcommand{\ie}{{\it i.e.,\ }}

\newcommand{\beq}{\begin{equation}}
\newcommand{\eeq}{\end{equation}}
\newcommand{\bea}{\begin{eqnarray}}
\newcommand{\eea}{\end{eqnarray}}
\newcommand{\ba}{\begin{array}}
\newcommand{\ea}{\end{array}}
\newcommand{\bi}{\begin{itemize}}
\newcommand{\ei}{\end{itemize}}
\newcommand{\bn}{\begin{enumerate}}
\newcommand{\en}{\end{enumerate}}
\newcommand{\bc}{\begin{center}}
\newcommand{\ec}{\end{center}}
\renewcommand{\l}{\left}
\renewcommand{\r}{\right}

\newcommand{\eq}[1]{Eq.~(\ref{#1})}
\newcommand{\eqs}[2]{Eqs.~(\ref{#1}) and (\ref{#2})}
\newcommand{\eqss}[3]{Eqs.~(\ref{#1}), (\ref{#2}) and (\ref{#3})}

\newcommand{\MeV}{\mathinner{\mathrm{MeV}}}
\newcommand{\GeV}{\mathinner{\mathrm{GeV}}}

\begin{document}

\preprint{FTUV-15-09-25}
\preprint{IFIC-15-50}

\title{On the tensor-to-scalar ratio in large single-field inflation models}


\author{Gabriela Barenboim}
\email[]{Gabriela.Barenboim@uv.es}
\author{Wan-Il Park}
\email[]{Wanil.Park@uv.es}
\affiliation{
Departament de F\'isica Te\`orica and IFIC, Universitat de Val\`encia-CSIC, E-46100, Burjassot, Spain}


\date{\today}

\begin{abstract}
We show that generically the tensor-to-scalar ratio in large single-field inflation scenarios is bounded to be larger than $\mathcal{O}(10^{-3})$ for the spectral index in the range favored by observations.
\end{abstract}


\maketitle


\section{Introduction}
In scenarios of inflation, a necessary step to explain various features of the present universe, the number of $e$-foldings ($N_e$) is conventionally taken to be around $50$-$60$ depending on the cosmological history after inflation.
The lower bound of $N_e$ can be pushed down as long as the reheating temperature after inflation is high enough to provide the relevant physics afterwards (e.g., baryogenesis, production of dark matter, a successful Big-Bang nucleosynthesis).
However, the upper bound is saturated at about $60$, since the upper bound of the tensor-to-scalar ratio ($r_T < 0.1$ \cite{Ade:2015xua}) coming from the non-observations of the tensor mode perturbations, puts an upper bound on the energy scale of inflation \ie $V_*^{1/4} \lesssim 2 \times 10^{16} \GeV$.

In conventional Einstein gravity, the tensor-to-scalar ratio is simply proportional to the slow-roll parameter associated with the slop of the inflaton potential.
For small field inflation in which the excursion scale of the inflaton can be lowered down far below Planck scale, $r_T$ can be extermely small.
However, in large-field inflation scenarios, this can not be the case since the $e$-foldings are likely to turn out too large unless the potential has an unphysical sudden change.
Hence, $r_T$ is likely to be lower-bounded in such scenarios.

In this work, we show that the $e$-foldings required to match observations set a lower bound of $r_T$ at around $\mathcal{O}(10^{-3})$ in most of realistic models of large single-field inflation.
Our result is consistent with a recent work, Ref.~\cite{Huang:2015xda}, although the approach is different.

\section{Inflation and the number of  $e$-foldings for a given scale}
In slow-roll inflation scenario with a potential $V$ of a single scalar field (playing the role of inflaton), the slow-roll parameters are defined as
\beq
\epsilon \equiv \frac{1}{2} \l| \frac{M_{\rm P} V'}{V} \r|^2, \quad \eta \equiv \frac{M_{\rm P}^2 V''}{V}, \quad \xi^2 \equiv \frac{M_{\rm P}^4 V' V''}{V^2}
\eeq
where $M_{\rm P}=2.4 \times 10^{18} \GeV$ is the reduced Planck mass, and `$\prime$' represents derivative with respect to the inflaton field.
The power spectrum of scalar and tensor mode perturbations are given by
\beq \label{PR-PT}
P_\mathcal{R} = \l( \frac{H}{2 \pi} \r)^2 \frac{1}{2 \epsilon M_{\rm P}^2}, \quad
P_T = \l( \frac{H}{2 \pi} \r)^2 \frac{8}{M_{\rm P}^2}
\eeq
respectively, giving a tensor-to-scalar ratio
\beq \label{rT}
r_T \equiv \frac{P_T}{P_\mathcal{R}} = 16 \epsilon
\eeq
The spectral indices of these modes are 
\beq \label{ns-nT}
n_s = 1 - 6 \epsilon + 2 \eta, \quad
n_T = - 2 \epsilon
\eeq
Note that $n_T=-r_T/8$, a consistency relation of inflation in Einstein gravity.
For inflation to end smoothly, $\epsilon$ should depend on the scale; although  $\eta$ can be constant. The spectral running, $\alpha \equiv d n_s/ d \ln k$, is given by 
\beq
\alpha = 16 \epsilon \eta - 24 \epsilon^2 - 2 \xi^2
\eeq

The number of $e$-foldings generated until the end of inflation, since a given cosmological scale (denoted with a subscript `$_*$') exits the horizon during inflation, is given by
\beq \label{Ne-th}
N_{e,*}^{\rm th} \equiv \int_*^e H dt \simeq - \frac{1}{M_{\rm P}^2} \int_*^e d \phi \frac{V}{V'}
\eeq
with $\phi$ being the inflaton field.
For a given model of inflation (i.e., a specific form of potential $V$), the observed $n_s$ fixes $\phi_*$ and hence $N_{e,*}^{\rm th}$ is also fixed.
Meanwhile, in order for inflation to solve the smoothness problem of Big-Bang cosmology, any two points within our Hubble patch had to be in casual contact at some stage before the end of inflation. That requires a minimum amount of $e$-foldings, that we will denote $N_{e,*}^{\rm obs}$.
In order for an inflation scenario to work, $N_{e,*}^{\rm th} = N_{e,*}^{\rm obs}$ is required, and this constrains the possible models of inflation.
If there were no more entropy production after the reheating following inflation, the $e$-foldings associated with the Planck pivot scale ($k_* = 0.05 {\rm Mpc}^{-1}$) are given by \cite{KT}
\bea \label{Ne-obs}
N_{e,*}^{\rm obs} 
&=& \frac{1}{3} \ln \l[ \frac{2 \sqrt{6} \pi^3}{3} \frac{s_0}{k_*^3} \l( \frac{V_*}{V_e} \frac{V_e^{\frac{1}{4}}}{M_{\rm P}} \r)^3 \l( \frac{T_{\rm d}}{V_e^{\frac{1}{4}}} \r) \r]
\nonumber \\
&\simeq& 57.87 - \frac{1}{3} \ln \l[ \l( \frac{V_e}{V_*} \r)^3 \l( \frac{10^{16} \GeV}{V_e^{\frac{1}{4}}} \r)^3 \frac{V_e^{\frac{1}{4}}}{T_{\rm d}} \r]
\eea
where $s_0$ is the present entropy density
\beq
s_0 = \frac{2 \pi^2}{45} g_{*s0} T_0^3
\eeq
with $g_{*s0} = 3.9091$ being the number of relativistic degrees of freedom and $T_0=2.726 {\rm K}$ the CMB temperature at the present universe. The subscript `$_e$' in the right-hand side denotes the value at the end of inflation, and $T_{\rm d}$ is the decay temperature of the inflaton.  
The current upper-bound on the tensor-to-scalar ratio $r_T < 0.1$ \cite{Ade:2015xua} implies 
\beq
V_*^{1/4} \lesssim 2 \times 10^{16} \GeV
\eeq
and the reheating temperature is bounded as 
\beq
1 \MeV \lesssim T_{\rm d} \lesssim V_*^{1/4}
\eeq
Then, setting $V_*=V_e$, one finds 
\beq
N_{e,*}^{\rm obs} \lesssim 58.56
\eeq
The lower bound of $N_{e,*}^{\rm obs}$ depends on the post-inflation cosmology as well as $V_*$ and $T_{\rm d}$.
In particular, there can be an $\mathcal{O}(10)$ additional contribution to the $e$-foldings of inflation from, for example, thermal inflation \cite{Lyth:1995hj,Lyth:1995ka}.
In this case, the required $e$-foldings from primordial inflation are reduced by that amount.
In principle, there can be multiple stages of thermal inflation, providing a few tens of $e$-foldings.
However, thermal inflation with more than two stages is non-trivial to realize (or not so realistic), and too much extra $e$-foldings may cause some inconsistency with observations of small scale structure (see for example \cite{Hong:2015oqa}).
In realistic models, we expect extra $e$-foldings of around $20$ or less \cite{Jeong:2004hy,Kim:2008yu,Choi:2009qd,Park:2010qd,Choi:2012ye}.
Hence, in this letter we take $\Delta N_e = 15$ as the plausible maximal extra $e$-foldings with $T_{\rm d} = V_*^{1/4}$ for simplicity in the case where thermal inflation is considered.

The number of $e$-foldings in \eq{Ne-th} can be separated into contributions from slow-roll and non-slow-roll (or fast-roll) regions, denoted respectively as $\Delta N_{e, \rm sr}$ and $\Delta N_{e, \rm fr}$, i.e.,
\beq
N_{e,*}^{\rm th} = \Delta N_{e, \rm sr} + \Delta N_{e, \rm fr} 
\eeq
where
\beq
N_{e, \rm sr} = \int_*^\times \frac{d \phi/M_{\rm P}}{\sqrt{2 \epsilon}}, \quad N_{e, \rm fr} = \int_\times^e \frac{d \phi/M_{\rm P}}{\sqrt{2\epsilon}}
\eeq
The value of $\phi_\times$ for which the slow-roll approximation breaks down depends on the potential and is rather ambiguous.
However, we can take $\epsilon_\times \equiv \epsilon(\phi_\times) \sim \mathcal{O}(0.1)$, and in this case   
\beq
N_{e, \rm fr} < \frac{\Delta \phi/M_{\rm P}}{\epsilon_\times} \simeq 2.24 \l( \frac{0.1}{\epsilon_\times} \r)^{1/2} \frac{\Delta \phi}{M_{\rm P}}
\eeq
where $\Delta \phi \equiv |\phi_e - \phi_\times|$ with $\phi_e$ being the value of the inflaton at the end of inflation.
In large-field models, generically, $\Delta \phi \gtrsim M_{\rm P} \sim 1$, being the  precise value model dependent.
Hence, as a simple conservative constraint, we require 
\beq \label{Ne-bnd}
\Delta N_{e, \rm sr} < N_{e,*}^{\rm th} - \Delta N_e
\eeq
with $\Delta N_e =0, 15$ depending on the existence of thermal inflation.

\section{Classes of potentials}
There are numerous different shapes of potentials for slow-roll inflation.
However, in the region of field values where the slow-roll approximation is still valid, inflaton potentials generally fall into one of the following forms,
\begin{itemize}
\item Chaotic monomial: $V_{\rm ch} = V_0 x^p$
\item Inverse-Hilltop: $V_{\rm iht} = V_0 \l( 1 - 1/x^p + \dots \r)$
\item Starobinsky-like: $V_{\rm st} = V_0 \l( 1 - e^{-x} + \dots \r)$
\item Hilltop: $V_{\rm ht} = V_0 \l( 1 - x^p + \dots \r)$
\end{itemize}
where $x = \phi/\mu$ with $\mu$ being a scale characterizing the end of inflation, and $p$ is assumed to be positive definite. 
Note that, even if the potentials listed above are approximated froms, it is clear that they all are significantly changed as $x \to 1$.
Hence, unless the potentials have non-trivial complications around the end of inflation, we expect that inflation ends at $x \sim 1$.
In the following subsections, we show the approximate forms of slow-roll parameters and $N_{e,*}^{\rm th}$ or $N_{e,\rm sr}$ (depending on its relevance) which are expected to be valid as long as $\epsilon$ and $\eta$ are much smaller than unity in the region where $V \approx V_0$. 
These approximate expressions are useful to get an idea of the parametric dependences of the relevant quantities. The discrepancies one would get when considereing a complete potential are minor and do not change our argument.

\subsection{Chaotic monomial}
In this case, the slow-roll parameters take the (exact) form
\beq
\epsilon = \frac{p^2}{2 x^2} \l( \frac{M_{\rm P}}{\mu} \r)^2, \quad \eta = \frac{p (p-1)}{x^2} \l( \frac{M_{\rm P}}{\mu} \r)^2
\eeq
giving $\eta=2(p-1) \epsilon/p$, and from \eq{ns-nT}
\beq \label{ep-star-Vch}
\epsilon_* = \l( \frac{p}{p+2} \r) \frac{1-n_s}{2}
\eeq 
which does not depend on $\mu$.
The $e$-foldings are
\beq
N_{e, *}^{\rm th} = \frac{p}{4} \l( \frac{1}{\epsilon_*} - 1 \r) \simeq \frac{p+2}{2 (1-n_s)}
\eeq
It is clear that, as $p$ becomes larger than 1, $\eta$ becomes larger than zero, and similarly $\epsilon$.
Also, only $p \lesssim 4$ is allowed, othewise too much $e$-foldings are expected.
As can be seen from \eq{ep-star-Vch}, $\epsilon$ can be lowered down by taking a small $p$.
However, in large field scenarios in which $x_* > 1$ with $\mu \geq M_{\rm P}$, we are constrained to have $p \gtrsim \mathcal{O}(10^{-2})$. 
Hence, $r_T$ is lower-bounded at $\mathcal{O}(10^{-3})$.

\subsection{Inverse-Hilltop potential}
For $V \approx V_0$, the slow-roll parameters can be approximated as 
\beq \label{Viht-s-para}
\epsilon \approx  \frac{p^2}{2 x^{2(p+1)}} \l( \frac{M_{\rm P}}{\mu} \r)^2, \quad \eta \approx -\frac{p (p+1)}{x^{p+2}} \l( \frac{M_{\rm P}}{\mu} \r)^2
\eeq
From \eq{Viht-s-para}, we find
\beq \label{eta-Viht}
\eta = - \frac{2(p+1)}{p} \l[ \frac{2}{p^2 (M_{\rm P}/\mu)^2} \r]^{q-1} \epsilon^q
\eeq
where 
\beq \label{q-Viht}
\frac{1}{2} <q\equiv\frac{p+2}{2(p+1)}<1
\eeq
Using \eq{ns-nT} with \eq{eta-Viht}, one can find $\epsilon_*$, at least numerically, as a function of $p$ and $n_s$ for a given $\mu$.
Note that $\eta/\epsilon = - 2 x^p (p+1)/p$ in this potential, which means that for a given $\eta$, as $p$ goes away from a value around one (or $\mu$ is lowered down), $\epsilon$ decreases.
The $e$-foldings for slow-roll regime are
\beq \label{Ne-Viht}
N_{e, \rm sr} \approx \frac{x_*^{p+2}-x_\times^{p+2}}{p (p+2)} \l( \frac{\mu}{M_{\rm P}} \r)^2 \approx \l( \frac{1}{\eta_\times} - \frac{1}{\eta_*} \r) \l(\frac{p+1}{p+2} \r)
\eeq
where $\eta_\times$ is found from \eq{eta-Viht} with $\epsilon_\times = 0.1$.

We can explore the limiting cases of $p$ leading to small $\epsilon_*$. When $p\to0$, one finds
\beq
\epsilon_* \simeq \frac{p}{4} \l( 1 - n_s \r), \ N_{e, \rm sr} \simeq \frac{1}{1-n_s}
\eeq
Hence, taking a small $p$, one can lower down $\epsilon_*$, but $N_{e, \rm sr} \gtrsim 50$ requires $n_s \gtrsim 0.98$ which is out of the 2-$\sigma$ region. 
On the other hand, if $p \to \infty$,
\beq \label{ep-star-Viht}
\epsilon_*^{1/2} \simeq - \frac{\gamma}{6} \l( 1 - \sqrt{1+ \frac{6 (1-n_s)}{\gamma^2}} \r), \ N_{e, \rm sr} \simeq \frac{2}{1-n_s}
\eeq
with $\gamma \equiv \sqrt{2} p (M_{\rm P}/\mu)$.
It may be natural and tempting to set $\mu$ to be the Planck scale. 
However, in principle $\mu$ can be much larger than $M_{\rm P}$ although it may need some fine-tuning or non-trivial (or unnatural) realizations. 

\subsection{Starobinsky-like}
In this case, one finds
\beq \label{Vst-s-para}
\epsilon \approx  \frac{e^{-2x}}{2} \l( \frac{M_{\rm P}}{\mu} \r)^2, \quad \eta \approx - e^{-x} \l( \frac{M_{\rm P}}{\mu} \r)^2
\eeq
Note that, as $\mu$ is increased, $\epsilon_*$ matching observations increases as well.
So, we take $\mu=M_{\rm P}$ to see the smallest allowed $\epsilon_*$.
From \eq{Vst-s-para}, one finds
\beq \label{eta-Vst}
\eta = - \sqrt{2} \epsilon^{1/2}
\eeq
leading to
\beq \label{ep-star-Vst}
\epsilon_*^{1/2} = - \frac{\sqrt{2}}{6} \l( 1 - \sqrt{1+ 3 (1-n_s)} \r)
\eeq
which is the same as $\epsilon_*^{1/2}$ in \eq{ep-star-Viht} with $\mu/M_{\rm P}=p$.
The the number of $e$-foldings for slow-roll regime are 
\beq
N_{e, \rm sr} \approx \frac{1}{\eta_\times} - \frac{1}{\eta_*}
\eeq
where $\eta_\times$ is obtained from \eq{eta-Vst} with $\epsilon_\times = 0.1$.

\subsection{Hilltop potential}
In this case, for $0<p<1$ inflaton should be located at a particular region in $0<x<1$ as the initial condition, and this is non-trivial to realize.
Also, as $p$ increases, the $\epsilon_*$ needed to match observation becomes smaller. 
So, we consider only $p\geq2$ in order to avoid irrelevant complications.
Slow-roll parameters are given as
\beq
\epsilon \approx \frac{p^2 x^{2(p-1)}}{2} \l(\frac{M_{\rm P}}{\mu} \r)^2, \quad \eta \approx -p (p-1) x^{p-2} \l(\frac{M_{\rm P}}{\mu} \r)^2
\eeq
Note that for $p=2$, $\eta$ is a constant depending on $\mu$ exclusively, and 
\beq
\frac{\mu}{M_{\rm P}} = \frac{2}{\sqrt{1-n_s-6\epsilon_*}} \geq \frac{2}{\sqrt{1-n_s}} \sim \mathcal{O}(10) 
\eeq
For $p \neq1,2$, $\eta$ can be found in terms of $\epsilon$ from \eqs{eta-Viht}{q-Viht} with $p \to -p$.
The number of $e$-foldings is
\beq
N_{e,\rm sr} \approx \l\{
\begin{array}{ll}
\frac{1}{2 \eta} \ln \l( \frac{\epsilon_\times}{\epsilon_*} \r) &: \ p=2
\\
\l( \frac{1}{\eta_\times} -\frac{1}{\eta_*} \r) \l( \frac{p-1}{p-2} \r) &: \ p \neq 1, 2
\end{array}
\r.
\eeq
Similarly to the case of $V_{\rm iht}$, as $\mu$ decreases, $\epsilon_*$ decreases, and we consider $\mu/M_{\rm P}\geq1$ here too.
Note that, if $\mu/M_{\rm P}=p$, as $p\to \infty$, $\epsilon_*$ collapses to the one in \eq{ep-star-Vst}.
%
\begin{figure*}[t]
\begin{center}
\includegraphics[width=0.32\textwidth]{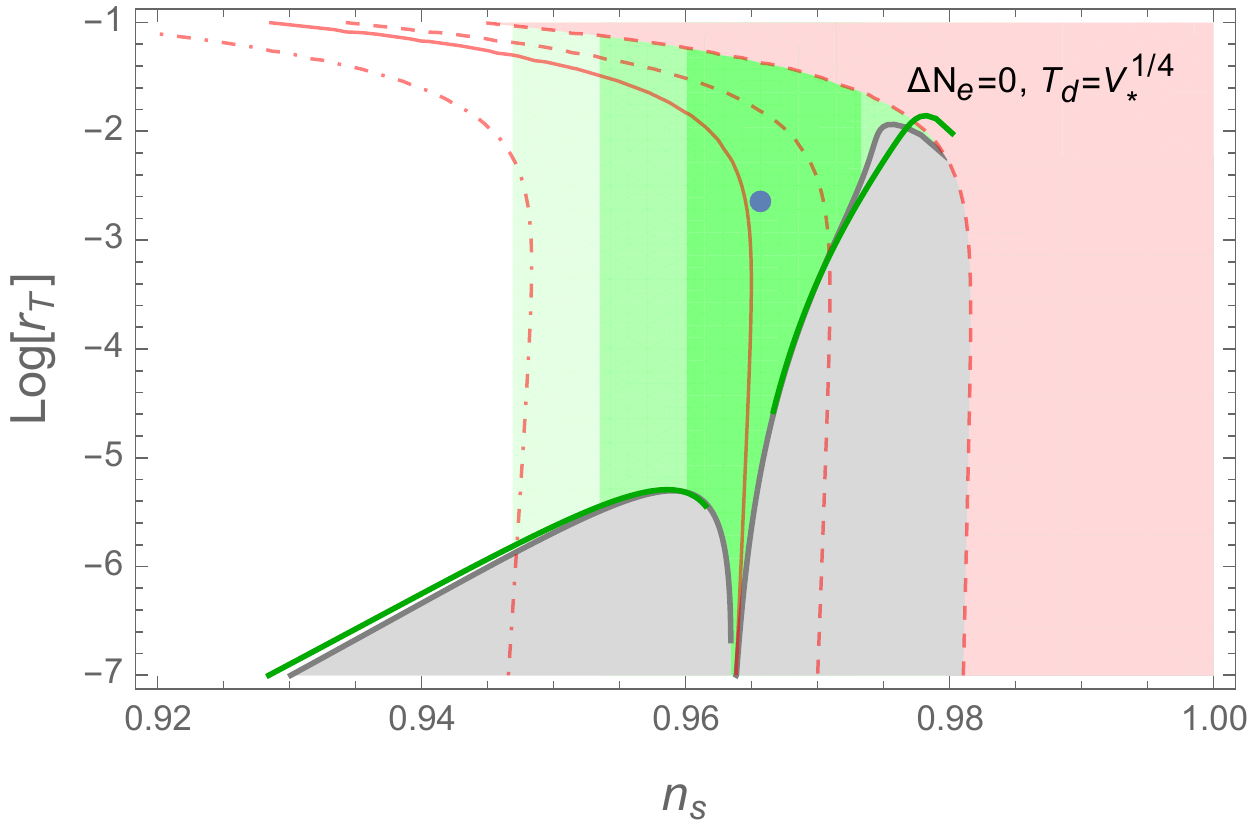}
\includegraphics[width=0.32\textwidth]{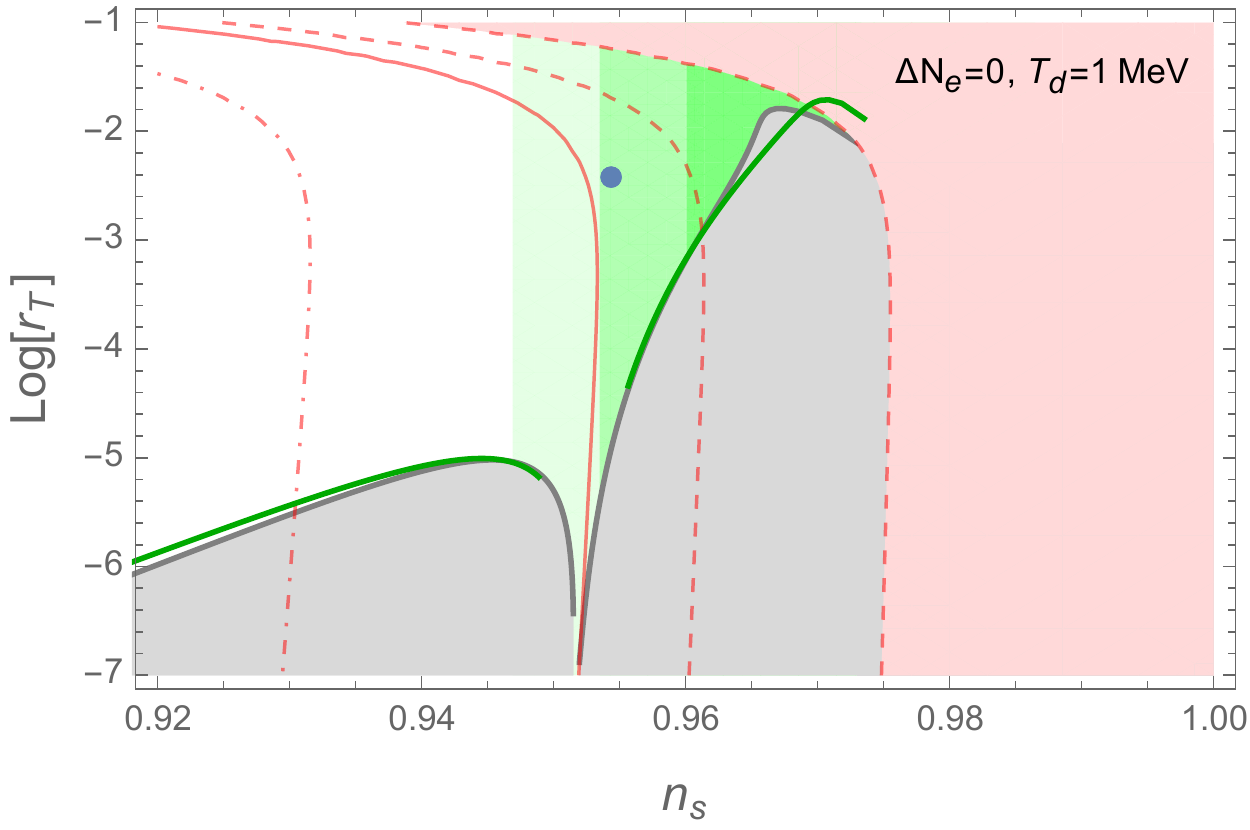}
\includegraphics[width=0.32\textwidth]{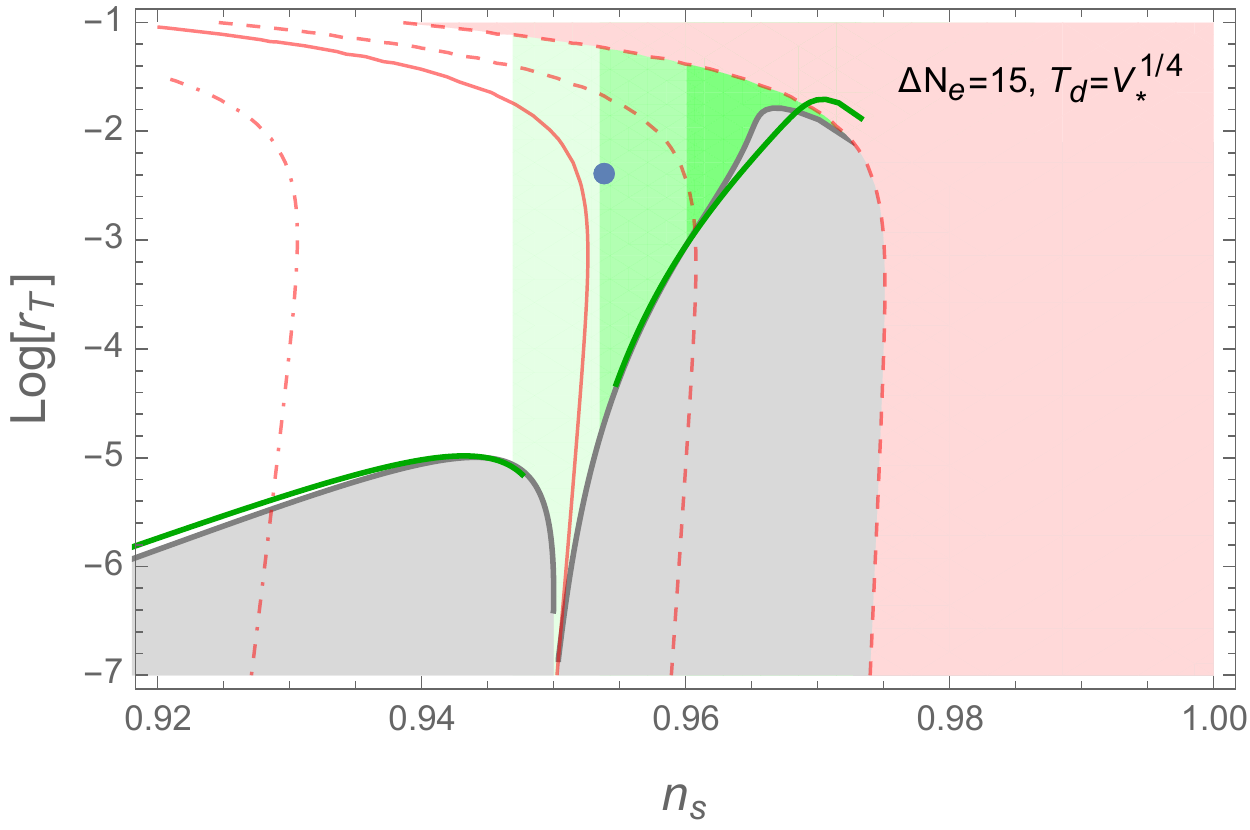}
\includegraphics[width=0.32\textwidth]{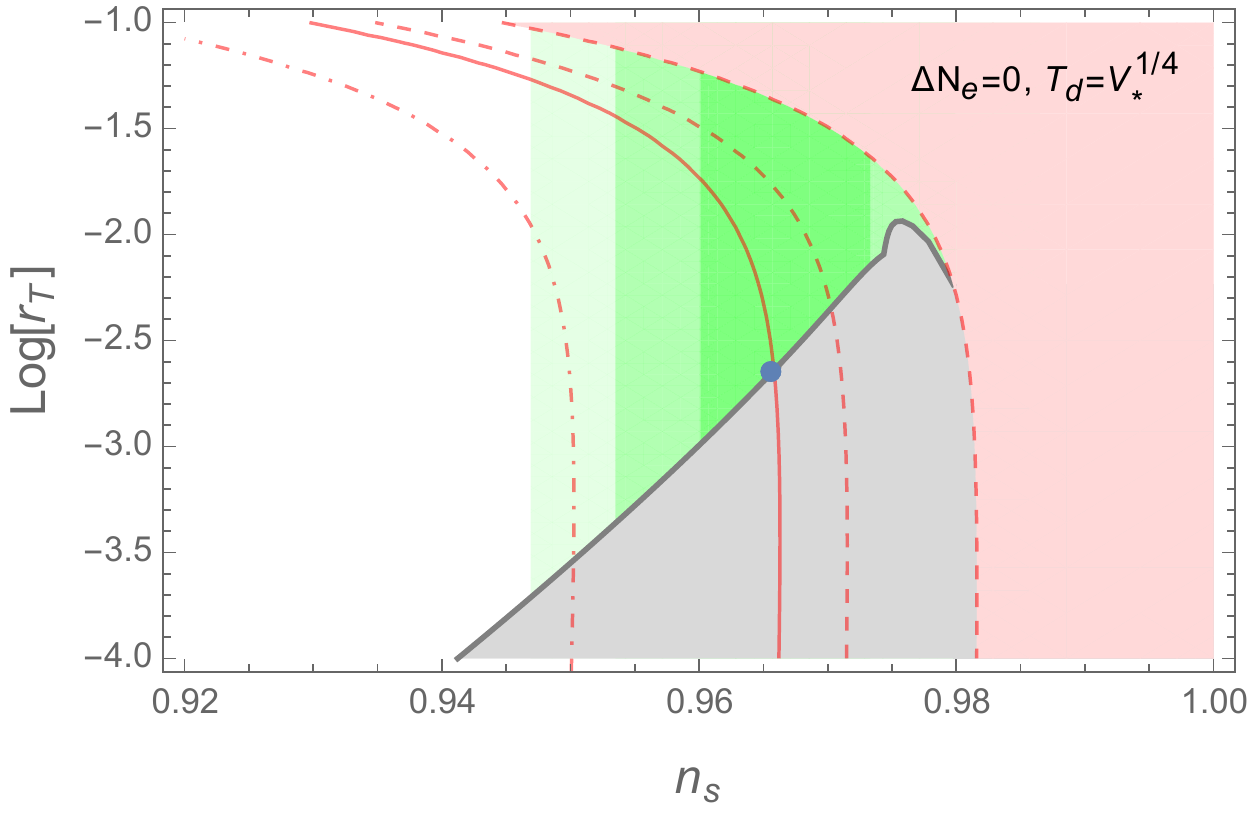}
\includegraphics[width=0.32\textwidth]{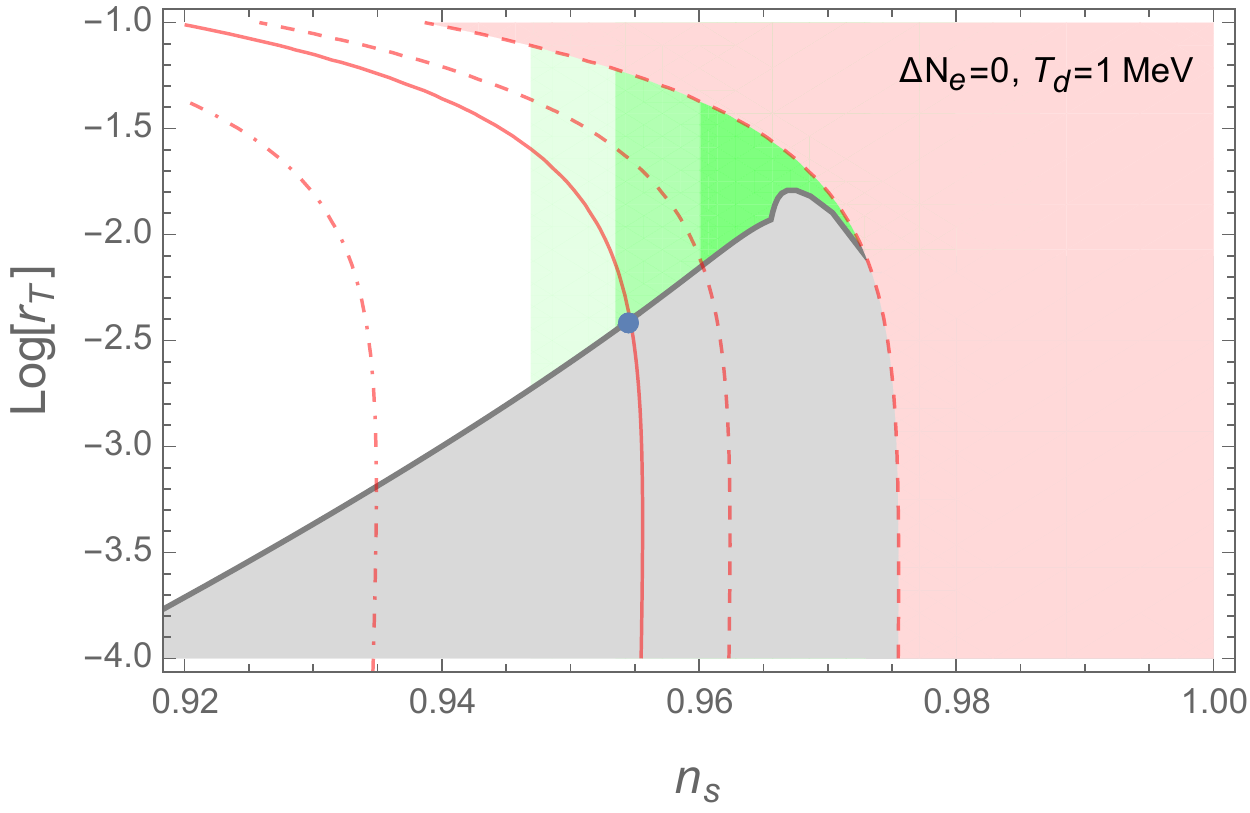}
\includegraphics[width=0.32\textwidth]{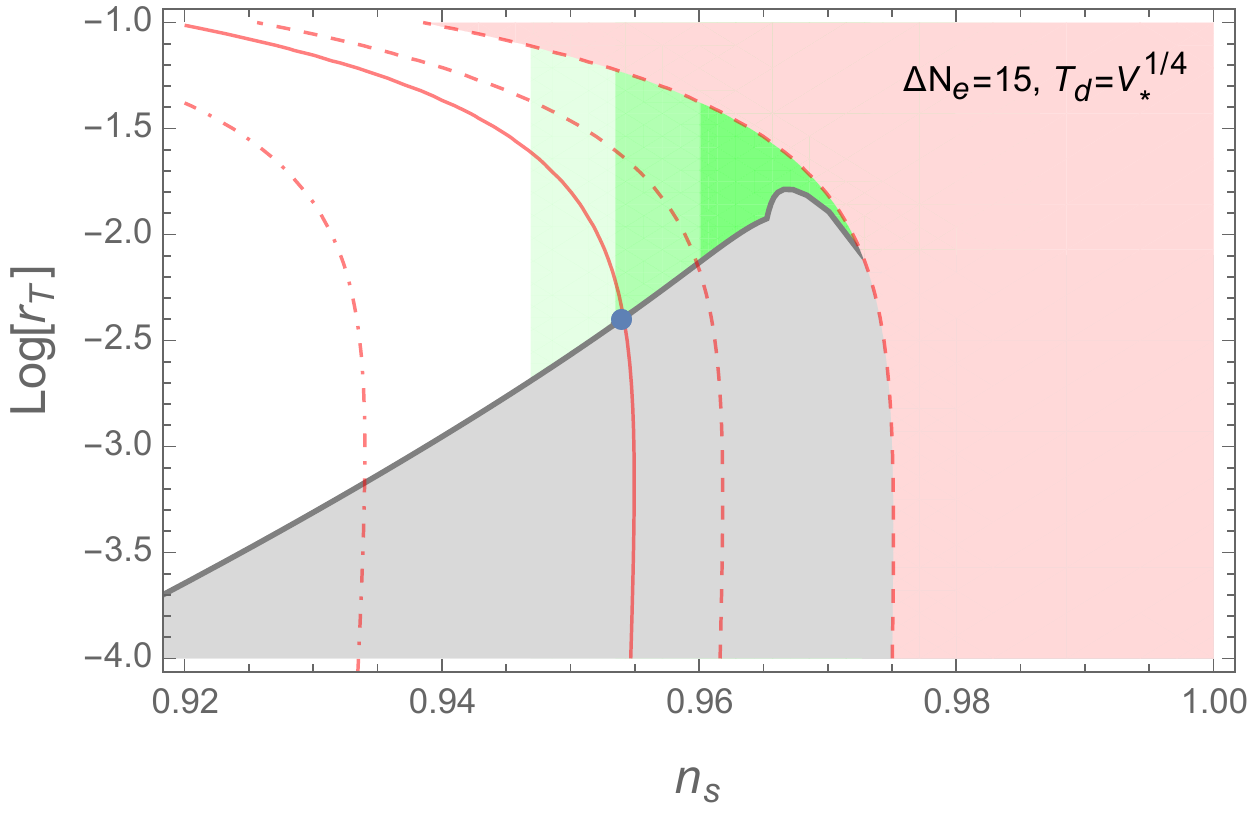}
\caption{The lower bound of $r_T$ (gray line(s)) as a function of $n_s$.
Green regions are 1, 2, 3-$\sigma$ bands of $n_s$ from Planck observations (Planck TT+lowP) \cite{Ade:2015lrj}. 
Red lines are obtained by imposing the constraint on the $e$-foldings, \eq{Ne-bnd}.
Red dashed lines are for inverse-Hilltop potential with $p=0.1,4$ from right to left.
Red dotdashed line is for Hilltop potential with $p=4$.
Red solid line is for $p\to \infty$ in both inverse-Hilltop and Hilltop potentials. 
For each value of $p$, the right-side of the red line is excluded because of too much $e$-foldings. 
Green line(s) is the bound obtained from the completions of potentials, \eqss{Viht-full}{Vht-full}{Vst-full}.
Blue dot is the prediction of Starobinsky-like potential with $\mu=M_{\rm P}$. 
\textit{Upper}: $\mu=M_{\rm P}$. 
\textit{Lower}: $\mu=p M_{\rm P}$ for $p\geq1$, but $\mu=M_{\rm P}$ for $p<1$.
}
\label{fig:rT-Lbnd}
\end{center}
\end{figure*}
%

\section{The lower bound of $r_T$: Numerical results}

In order to find a lower bound of $r_T$, we used the expressions obtained in the previous section and performed a numerical analysis.
Also, for comparsion, we used the following completions of potentials:
\bea \label{Viht-full}
V_{\rm iht} &=& V_0 \l( 1 - 1/2 x^p \r)^2
\\ \label{Vst-full}
V_{\rm st} &=& V_0 \l( 1 - e^{-x}/2 \r)^2
\\ \label{Vht-full}
V_{\rm ht} &=& V_0 \l( 1 -  x^p/2 \r)^2
\eea
The result is shown in Fig.~\ref{fig:rT-Lbnd}.
We found that, if $\mu \sim M_{\rm P}$ (upper panels), when $p \gg 4$ in inverse-Hilltop and Hilltop potentials, it is possible to lower-down $r_T$ by many orders of magnitude relative to the current upper bound.
However, for $p\leq4$ which is likely to be the case, in the region of interest the field value is far away from the  end point of inflation, and  either $r_T \gtrsim \mathcal{O}(10^{-3})$ or $n_s$ is out of the 3-$\sigma$ bound of observations (in Hilltop).
As shown in the lower panels of Fig.~\ref{fig:rT-Lbnd}, if we take $\mu$ larger than $M_{\rm P}$, the case of $p\leq4$ in Hilltop potential can be within the preferred region of observations, but $r_T$ is pushed up to $\mathcal{O}(10^{-3})$.
So, we can conlude that, including chaotic monomials and Starobinsky-like potentials, in realistic models (probably) with $p\leq4$, $r_T$ is lower-bounded at about $10^{-3}$ or $n_s$ is out-of the 3-$\sigma$ allowed band. 
Fig.~\ref{fig:rT-Lbnd} also shows that, if the decay temperature of the inflaton is low or there is an extra contribution to $e$-foldings, the lower-bound of $r_T$ is pushed up.

\section{Conclusions}

In this paper, we examined the lower bound of the tensor-to-scalar ratio in large single-field scenarios of inflation.
For inflaton field values associated to the relevant cosmological scales, the inflaton potential may be approximated to either chaotic monomial, inverse-Hilltop, Hilltop, or Starobinsky-like potentials in which the leading field-dependent term is $(\phi/\mu)^{\pm p}$ with $0<p \leq 4$ or $e^{-\phi/\mu}$.
We showed that, if the dimensionful scale $\mu$ characterizing the end of inflation is Planck scale or larger, which is the case of large single-field scenarios, the tensor-to-scalar ratio is lower-bounded at $\mathcal{O}(10^{-3})$ for the range of the spectral index favored by observations.
Therefore, even if it will not be done in the near future, most large single-field inflation models will be probed as experiments reach a $r_T$ at the level of $10^{-3}$.

\section{Acknowledgements}
The authors acknowledge support from the MEC and FEDER (EC) Grants SEV-2014-0398 and FPA2014-54459 and the Generalitat Valenciana under grant PROME- TEOII/2013/017.
G.B. acknowledges partial support from the European Union FP7 ITN INVISIBLES (Marie Curie Actions, PITN-GA-2011-289442).

\end{document}